\documentclass[preprint2]{aastex}

\begin{document}
\title{Magnetic Equilibrium} 
\author{Andrei Gruzinov}
\affil{CCPP, Physics Department, New York University, 4 Washington Place, New York, NY 10003}

\begin{abstract}

We propose that generic magnetic equilibrium of an ideally conducting fluid contains a volume-filling set of singular current layers. 

Singular current layers should exist inside neutron stars. Residual dissipation in the singular current layers might be the main mechanism for the magnetic field decay. The slow decay of the field might be the clock responsible for triggering the magnetar flares.

~
~

\end{abstract}

\section{The magnetic equilibrium problem}

An equilibrium (no convection) star might still be a good conductor and contain magnetic fields -- with the magnetic stress balanced by the fluid pressure and gravity. We want to describe the generic features of such equilibrium magnetic fields.

This problem was clearly formulated many years ago by Chandrasekhar and Fermi, but it remains unsolved to this day. Recent progress came from  numerical work (Braithwaite \& Nordlund 2005).  

The recent numerical work does not point out an important generic feature of the equilibrium --  singular current layers. Moreover, we propose that generic equilibrium carries infinitely many singular current layers. 

This proposition -- Generic equilibrium carries a volume-filling set of singular current layers -- is our main thesis. This is indeed just a proposition. It is  based on very shaky arguments. A numerical proof or refutation is needed.

The author tried to solve the problem numerically, but succeeded only in 2D, where the singular current lines are isolated rather than area-filling. Still,  it appears that modern computers, if handled by professionals, are powerful enough to also solve the 3D problem (by the relaxation technique, \S2).

 Since it is believed that close to equilibrium magnetic fields do exist in neutron stars and in other non-convective stars, a careful numerical study of the singularities should be worthwhile. If singularities are indeed generic, they might be responsible for some of the observed phenomena -- say, magnetic field decay and triggering the magnetar flares.

The singular current layer is a local feature, it can be considered without gravity. To simplify the analysis, we will also assume an incompressible constant density fluid. Then the magnetic fields ${\bf B}$ will be in equilibrium if the magnetic ${\bf j}\times {\bf B}$ force is balanced by the pressure gardient:
\begin{equation}\label{equi}
{\bf B}\times \nabla \times {\bf B}=-\nabla p. 
\end{equation}
We merely want to discuss solutions of eq.(\ref{equi}). 

\section{The relaxation method}

A well-known procedure for finding stable magnetic equilibria is the relaxation method. Consider a viscous but ideally conducting incompressible fluid. Set an arbitrary initial magnetic field and velocity, and let the system evolve. The total energy will decrease due to internal friction, and asymptotically the fluid must stop. The fluid can stop only in a magnetic equilibrium state. Since the fluid is an ideal conductor, the magnetic field topology is not changed by the fluid motion. Conserved topology plus incompressibility guarantee  that, generically, the equilibrium state reached by the fluid will be non-trivial, ${\bf B}\neq 0$. 

The relaxation method  proves the existence of magnetic equilibria with arbitrary topology. The method is constructive -- it  allows one to actually find this equilibrium. 

The corresponding mathematical problem, in the limit of large viscosity, and after an obvious generalization reads:
\begin{equation}\label{relax}
\begin{array}{c}
\dot{{\bf B}}=\nabla \times ({\bf v}\times {\bf B}),\\ 
{\bf v}=-\hat{L}\nabla \times \nabla \times ({\bf B}\times \nabla \times {\bf B}),
\end{array}
\end{equation}
where $\hat{L}$ is an arbitrary positive operator which commutes with $\nabla$.  The first equation says that the magnetic field ${\bf B}$ is frozen into the fluid moving with velocity ${\bf v}$. The second equation says that the fluid incompressibly yields to the magnetic force; $\hat{L}=\Delta ^{-2}$ corresponds to internal friction,  $\hat{L}=-\Delta ^{-1}$ corresponds to external friction ($\Delta \equiv \nabla ^2$).  

Eq.(\ref{relax}) describes a dissipative evolution. Magnetic energy 
\begin{equation}\label{energy}
W={1\over 2}\int d^3r B^2
\end{equation}
decreases:
\begin{equation}\label{energydot}
\begin{array}{c}
\dot{W}=-\int d^3r \left( \nabla \times ({\bf B}\times \nabla \times {\bf B})\right)\\
 \cdot \hat{L}\left( \nabla \times ({\bf B}\times \nabla \times {\bf B})\right) \leq 0 .
\end{array}
\end{equation}
The evolution stops only when 
\begin{equation}\label{equi2}
\nabla \times ({\bf B}\times \nabla \times {\bf B})=0, 
\end{equation}
which is equivalent to eq.(\ref{equi}).

We must conclude that for any solenoidal field, there  exists an incompressible deformation  of this field which is an equilibrium.

\section{The Arnold argument}
The following argument, due to Arnold (1986), explains why the magnetic equilibrium problem remains unsolved, despite the clear and simple logic of the relaxation method.

From eq.(\ref{equi}), ${\bf B}\cdot \nabla p=0$. This means that magnetic field lines lie on the surfaces $p$=const. But the magnetic field we started with in the relaxation method generically does not possess magnetic surfaces -- generic field lines fill up volumes, not surfaces. Clearly, the isotopological relaxation (\ref{relax}) cannot comb the intertwined volume-filling lines onto the nested tori $p$=const. 

And yet, the relaxation argument of $\S2$ is still valid. For any solenoidal field there must exist  at least one incompressible deformation of the field which is an equilibrium, because asymptotically, the motion described by eq.(\ref{relax}) stops. It also appears likely that for suitable operators $\hat{L}$, the relaxing field remains non-singular at any finite time, meaning that the relaxation method indeed stops at a well-defined asymptotic magnetic field (the final field may be singular and different for different operators $\hat{L}$).

\begin{figure}
\plottwo{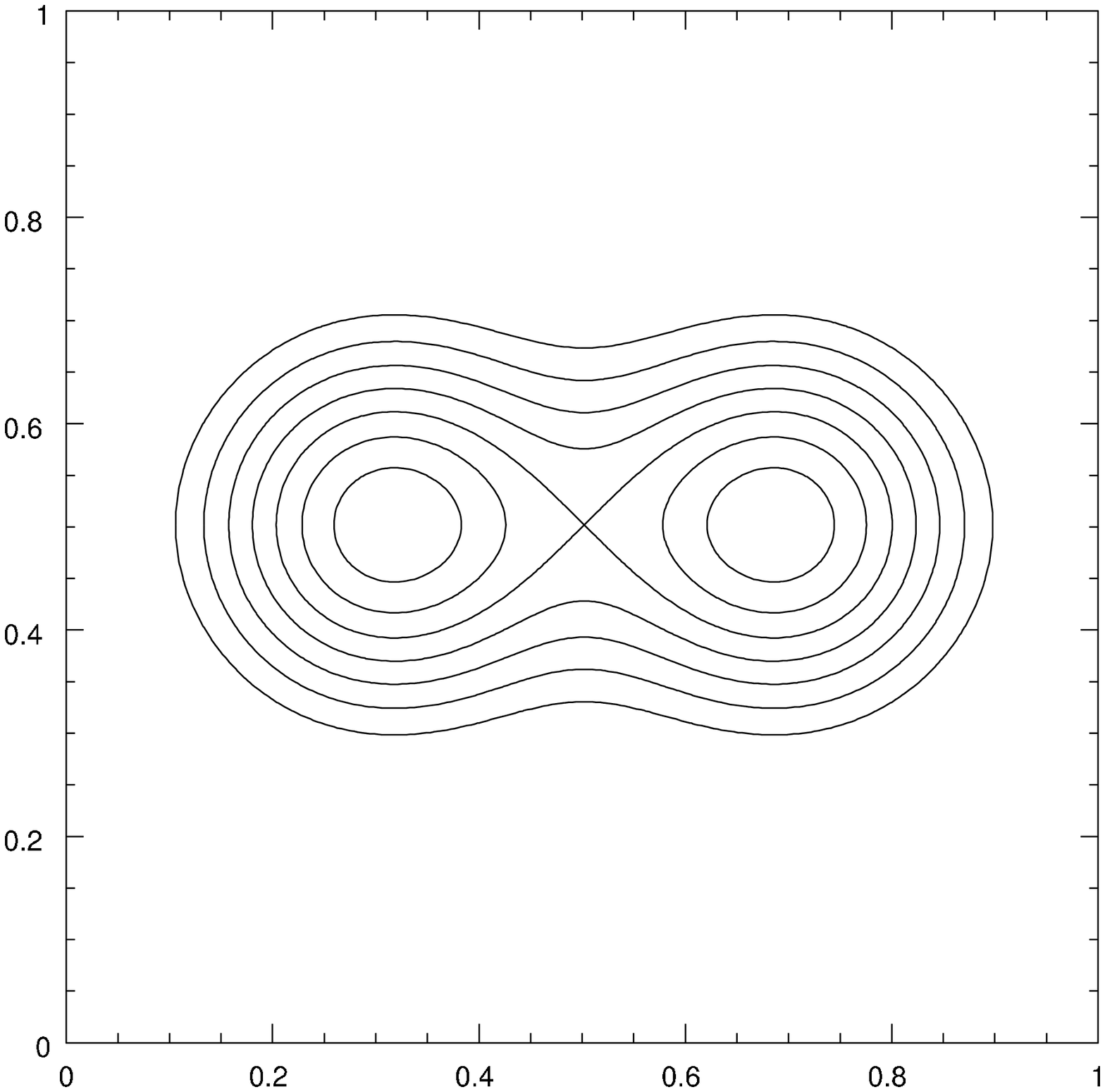}{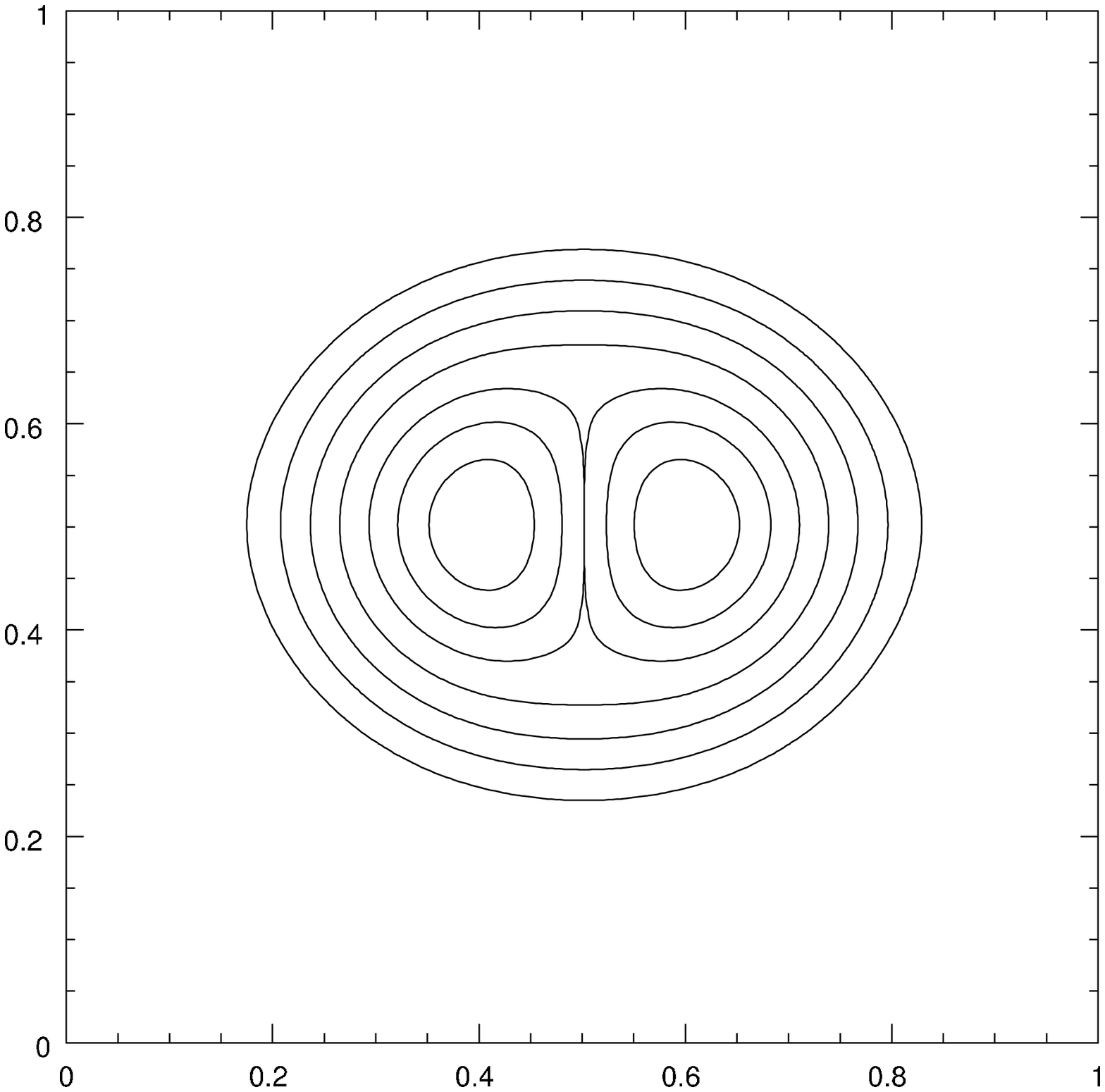}
\caption{{\it Left}: Initial isolines of $\psi$. {\it Right}: Final isolines of $\psi$. \label{fig1}}
\end{figure}

\begin{figure}
\plottwo{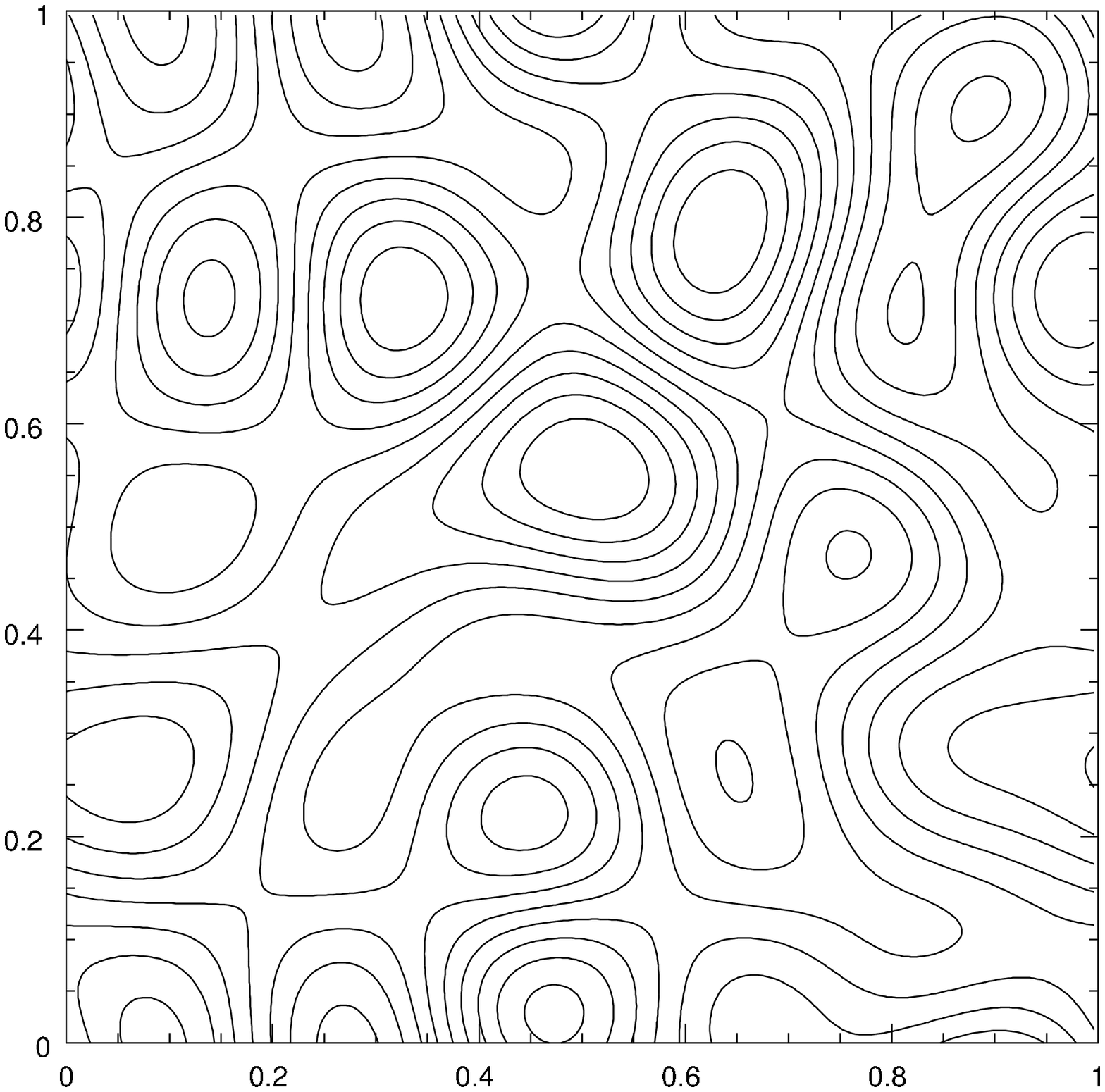}{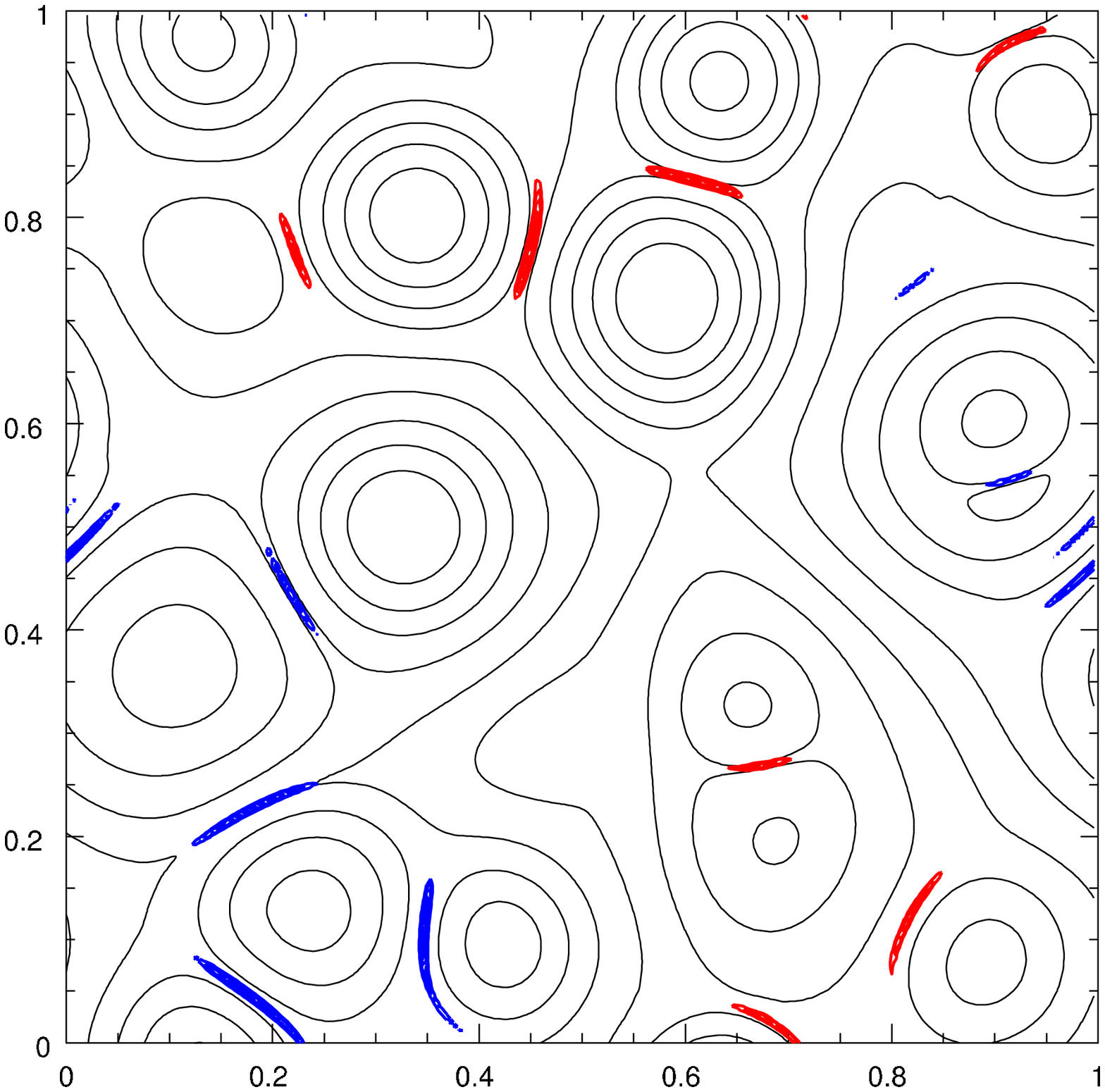}
\caption{{\it Left}: Initial isolines of $\psi$. {\it Right}: Final isolines of $\psi$, also shown (thick colored lines) are the isolines of $\Delta \psi$. \label{fig2}}
\label{fig2}
\end{figure}

\begin{figure} 
\includegraphics[angle=270, width=8cm]{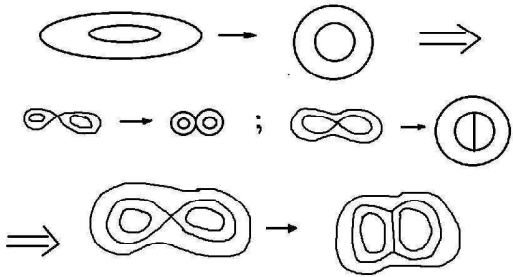}
\caption{Graphical "proof" that singular current layers form from the X-points}
\label{fig3}
\end{figure}

\section{ The 2D relaxation and equilibria }

After the above discussion, one really wants  to offer this problem to a computer and see how the computer will resolve this contradiction between topology and equilibrium. The author was able to obtain trustworthy numerical results only in 2D. In fact,  the 2D case is reach enough to show an important generic feature of magnetic equilibrium --  singular current layers. We discuss the 2D equilibria in this section.

First of all, it appears that in 2D there are no topological obstacles to equilibrium. Indeed, Arnold's argument says that magnetic field lines lie on the lines of constant pressure. But the 2D field lines are always the isolines of the magnetic stream function:  $\nabla \cdot {\bf B}$ gives ${\bf B}=(-\partial _y\psi, \partial _x \psi )$. 

Therefore one expects that a smooth equilibrium state will correspond to any initial state $\psi$. In reality, it has been known for some time that initial states with non-trivial topology relax to equilibrium states with singular current layers.\footnote{The author does not know who discovered that singular current layers are generic;  would be grateful for a reference.}

\begin{figure}
\plottwo{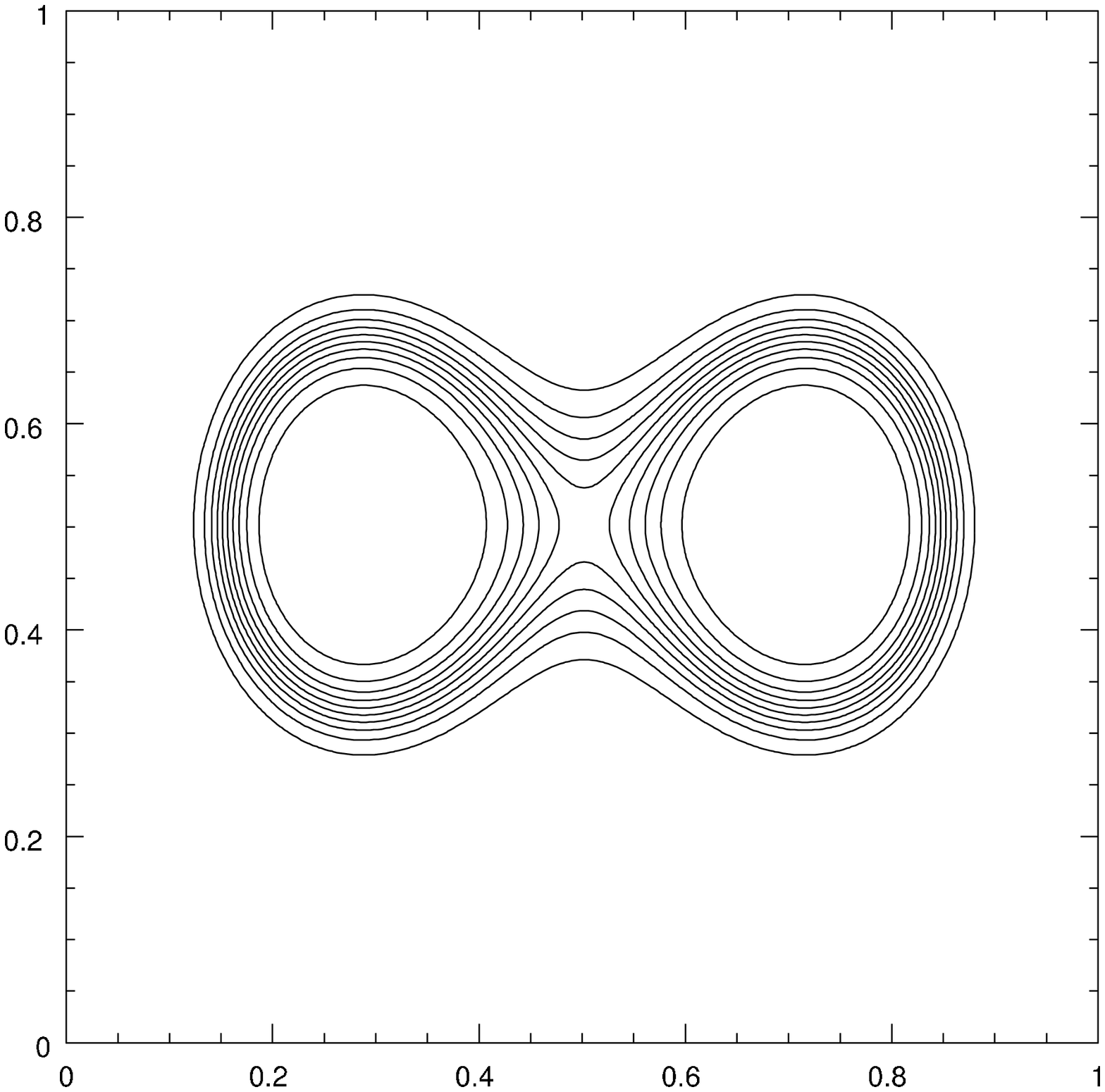}{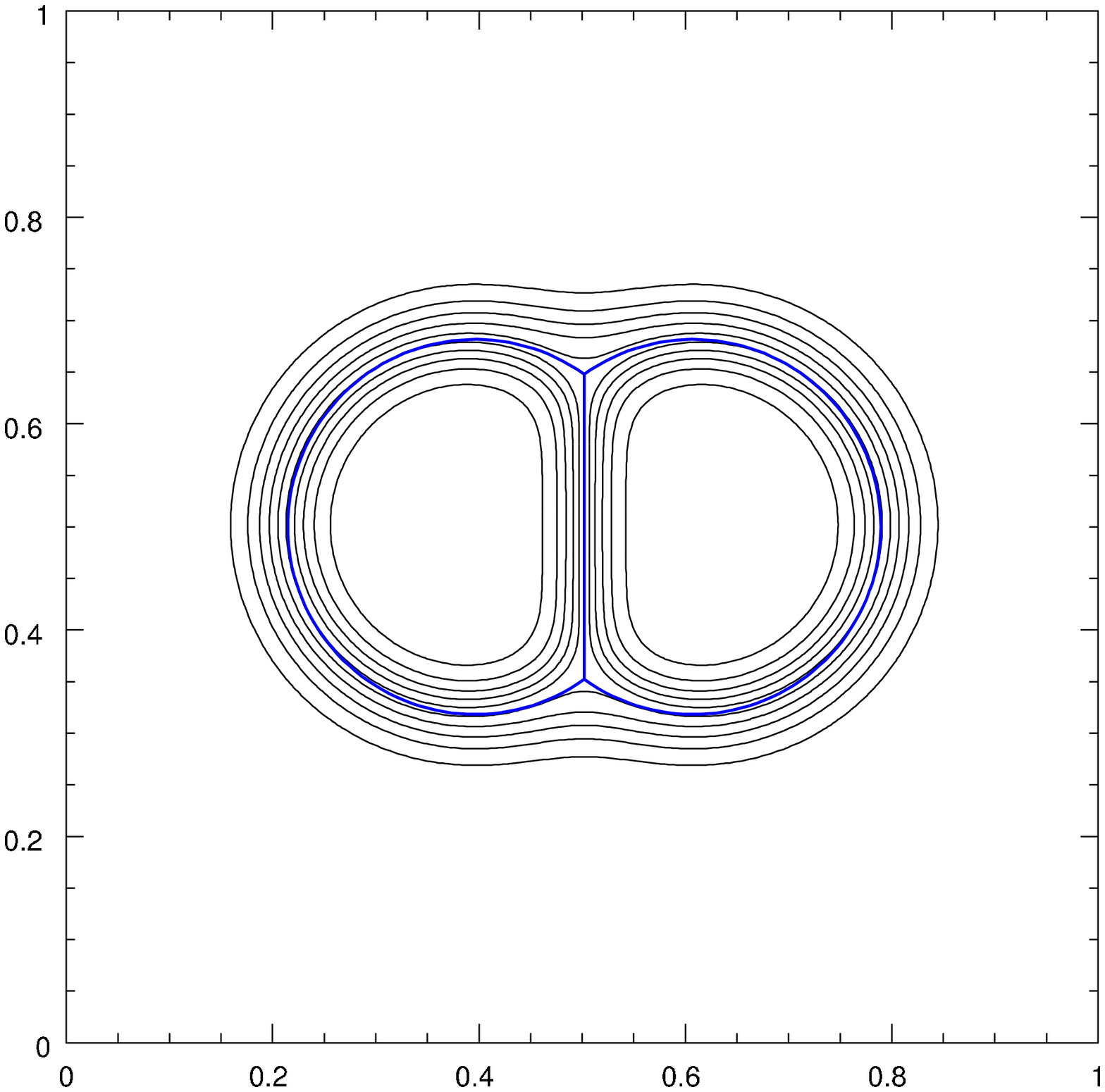}
\caption{{\it Left}: Initial isolines of $\psi$. {\it Right}: Final isolines of $\psi$. Also shown (thick colored line) is the theoretical separatrix -- two $251^\circ$ circular arcs and a line segment. }
\label{fig4}
\end{figure}

To show how the singular current layers form, consider relaxation with internal friction.  $\hat{L}=\Delta ^{-2}$ in eq.(\ref{relax}) gives the "incompressible diffusion" equation in 2D:
\begin{equation}\label{relax2}
\begin{array}{c}
\dot{\psi}=\{ \chi ,\psi \},~~~ \Delta ^2\chi =\{ \psi , \Delta \psi \},\\
\{ f, g\} \equiv \partial _xf\partial _yg-\partial _yf\partial _xg.
\end{array}
\end{equation}
The magnetic energy decreases
\begin{equation}\label{relax2W}
W={1\over 2}\int d^2r (\nabla \psi)^2,~~~\dot{W}=-\int d^2r(\Delta \chi )^2.
\end{equation}
The evolution stops at equilibrium
\begin{equation}\label{equi2}
\{ \psi , \Delta \psi \}=0,
\end{equation}
or, in another form, 
\begin{equation}\label{equi21}
\Delta \psi =F(\psi ).
\end{equation}

Computer simulations of eq.(\ref{relax2}) (using fast Fourier transform) show that all initial saddles of $\psi$ develop into singular current layers, see figures (1) and (2). The singularities are of the type $\psi \propto |x|$, giving the field $B\propto {\rm sign} (x)$, and the current $j\propto \delta (x)$ -- singular current layers.

The reason for existence of  singular current layers  in 2D appears to be clear. The incompressible relaxation process (\ref{relax2}) tries to make the field lines as short as possible, while keeping the area enclosed by each field line constant. The field lines try to deform into circles. This is impossible for non-trivial initial topology. Then, as illustrated by fig.(\ref{fig3}), one does expect that the saddles (X-points) of the initial field  split into two Y-points with a singular current layer between them. \footnote{ Does the outer part of the separatrix become a singular current layer?  In the past, the author claimed that it does but the argument given was incorrect. Our numerical results are inconclusive.}

One can actually transform the cartoon fig.(\ref{fig3}) into a rigorous calculation. Take the limit in which only an infinitesimal layer of the field surrounds the separatrix. Assume  equal  initial energies in the outer and inner parts of the separatrix. Then, the magnetic energy will scale as $W\propto L_i^2+L_o^2$, where $L_i$ is the inner and $L_o$ is the outer lengths of the separatrix. Minimization of the energy at fixed enclosed areas predicts the final shape of the separatrix -- two $\approx 251^\circ$ circular arcs and a line segment separating the arcs, for a symmetric separatrix.  The direct numerical simulation confirms this result, fig.(\ref{fig4}).

\section{Perturbation theory in 2D and in 3D}

We have also performed 3D numerical simulations of the incompressible relaxation (\ref{relax}) with the inner friction, $\hat{L}=\Delta ^{-2}$. We see clear evidence for the singular current layers formation \footnote{ Numerical 3D singular current layers have been known for many years. } but we were unable to see how many singular layers form, how are they distributed in the volume, and how exactly does the system manage to preserve the field line topology, while putting the field lines onto the constant pressure surfaces. One probably needs a  better than our 64$^3$ resolution, and/or  a better handling of the numerical issues.
 
Having failed numerically, we resort to linear perturbation theory.  We start with a translation invariant equilibrium field, put a small perturbation on it, and ask what does the perturbed field relax into. 

While this linear problem can indeed be solved analytically, it is unclear what does the solution have to do with the generic equilibrium, which, of course, is fully nonlinear. We will see, however, that in 2D the linear perturbation theory can be used as a sensible guide to the properties of the generic equilibrium.  We will assume that the same is true in 3D, and this will allow us to propose that the generic magnetic equilibrium carries a volume-filling set of singular current layers. This kind of reasoning is unsatisfactory. But this all we can offer.

\subsection{Perturbation theory in 2D}

In 2D, the initial y-translation invariant background field is $\psi _0(x)$, corresponding to the magnetic field $B(x)=\psi _0'(x)$ in the y-direction. The perturbation is $\propto e^{iky}$, and we assume that $k\neq 0$ -- otherwise the perturbation simply re-difines the background field $\psi _0(x)$.   The final relaxed state will be $\psi _0(x)+\psi (x)e^{iky}$. To linear order, the equilibrium condition (\ref{equi2}) gives 
\begin{equation}\label{pert2}
B\psi ''-B''\psi -k^2B\psi =0.
\end{equation}
If $B\neq 0$ everywhere, one can write the perturbation as $\psi =B\phi$. Then
\begin{equation}\label{pert21}
(B^2\phi ')'-k^2B^2\phi =0.
\end{equation}
With the periodic or decreasing x-boundary conditions, this gives
\begin{equation}\label{pert20}
\int dx B^2(\phi '^2+k^2\phi ^2)=0,
\end{equation}
and $\phi =0$. This result is trivial -- just says that the field lines straighten up during the relaxation. 

Now consider the case when the background magnetic field $B$ has zeros.  In the short wavelength  limit, for $B(x_0)=0$, eq.(\ref{pert2}) reads
\begin{equation}\label{pert2u}
(x-x_0)(\psi '' -k^2\psi ) =0.
\end{equation}
The solution is 
\begin{equation}\label{pert2us}
\begin{array}{c}
\psi ''-k^2\psi \propto \delta (x-x_0),\\
\psi \propto e^{-k|x-x_0|},~~~k>0.
\end{array}
\end{equation}
This is a current singularity.

So far everything was fully rigorous. What is shaky, is our extrapolation of this result to the non-linear regime. Based on the linear perturbation theory, we "conclude" that the generic 2D equilibrium will have isolated singular layers at magnetic field zeros. This is indeed correct in 2D, if by zeros we mean only the saddle-point zeros.

\subsection{Perturbation theory in 3D}
In 3D, we choose a background field of the form ${\bf B}_0=(0, U(x), V(x))$. The perturbation is $\delta {\bf B}=(b_x, b_y, b_z)\propto e^{imy+ikz}$. Then the linearized equilibrium condition (\ref{equi}) is
\begin{equation}\label{pert31}
i(mU+kV)b_x-(Ub_y+Vb_z)'=p',
\end{equation}
\begin{equation}\label{pert32}
iV(kb_y-mb_z)+U'b_x=imp,
\end{equation}
\begin{equation}\label{pert33}
-iU(kb_y-mb_z)+V'b_x=ikp.
\end{equation}
Here $'\equiv d/dx$, and $p$ is the pressure perturbation. The magnetic field perturbation  is solenoidal 
\begin{equation}\label{pert34}
b_x'+imb_y+ikb_z=0.
\end{equation}

From (\ref{pert32}) and (\ref{pert34}) we calculate the product $Vb_z$ in terms of $b_x$ and $p$.  From (\ref{pert33}) and (\ref{pert34}) we calculate the product $Ub_y$ in terms of $b_x$ and $p$.  Plug these into (\ref{pert31}). The pressure perturbation $p$ cancels, and we get 
\begin{equation}\label{pert3f}
\begin{array}{c}
Wb_x''-W''b_x-(m^2+k^2)Wb_x=0, \\
W(x)\equiv mU(x)+kV(x)
\end{array}
\end{equation}
This equation is isomorphic to eq.(\ref{pert2}), and we can use the 2D results. If $W(x)$ has no zeros, the perturbation is zero. For the relaxation problem it means that the lines straighten up. But if $W$ does have zeros, each zero will develop into a singular layer.  

The difference between 2D and 3D is that in 3D the zeros of $W(x)$ depend on wavenumbers. Suppose we work in a $2\pi$-periodic cube. Then the wavenumbers $m$, $k$ are arbitrary  integers, and all points $x$ where $U(x)/V(x)$ is a rational number are zeros of $W$ for the appropriate choice of $m$ and $k$. 

"Based" on the above, we propose that generic magnetic equilibrium contains a volume-filling (meaning everywhere dense within some domains) set of singular current layers. 

This proposition leaves many questions unanswered:  
\begin{itemize}

\item Is this proposition  true at all? 

And assuming it is true:

\item How exactly does this infinite set of singular layers help resolve the topological obstacles to relaxation?

\item Where do the layers form? \footnote{ In the perturbative calculation, the singular layers form at rational $U/V$, that is on closed field lines.}

\item Which current layers, a few strongest ones, or a lot of weaker ones, are responsible for the residual damping, once a small (probably anomalous in the real world) resistivity is added to the problem?

\end{itemize}
 
\section{Summary}

We propose that generic magnetic equilibrium contains a volume-filling set of singular current layers, because topological obstacles to relaxation are everywhere dense. This is just a proposition. The proposition leaves many questions unanswered. The proposition might turn out to be completely false.

It would be interesting to see what a generic field relaxes into in 3D.

\acknowledgements

This work was supported by the David and Lucile Packard foundation.

\end{document}